\documentclass[12pt]{iopart}

\usepackage{mathrsfs}
\usepackage{indentfirst}
\usepackage[pdftex]{color,graphicx}
\usepackage[pdftex,bookmarks,unicode,colorlinks]{hyperref}
\usepackage{enumerate} % 罗列格式
\usepackage[numbers]{natbib} % 参考文献引用格式
\usepackage{yfonts} % Fraktur/Gothic fonts
\usepackage{stmaryrd} % 三条竖杠范数
\usepackage[title]{appendix} % 附录添加  Appendix
\usepackage{supertabular} % 允许表格分页
\usepackage{caption} % 图片标题
\usepackage{tikz-cd}
\usetikzlibrary{matrix,arrows,decorations.pathmorphing}

\makeatletter
\let\@fnsymbol\@arabic
\makeatother

\newtheorem{theorem}{Theorem}[section]
\newtheorem{remark}[theorem]{\underline{Remark}}

\newcommand{\E}{\mathbf E}
\renewcommand{\P}{\mathbf P}

\newcommand{\R}{\mathbf R}

\renewcommand{\S}{\mathbb S}

\newcommand{\D}{\mathcal D}

\renewcommand{\L}{\mathcal L}
\renewcommand{\e}{\epsilon}

\newcommand{\Pred}{\mathcal P}
\newcommand{\ind}{\mathbf 1}

\newcommand{\vf}[1]{\frac{\partial}{\partial{#1}}}
\renewcommand{\pt}{\partial}

\renewcommand{\D}{\mathbf{D}}

\newcommand{\divg}{\mathrm{div}\,}
\newcommand{\ts}[1]{\textstyle{{#1}}}

\begin{document}

\title[Stochastic Geometric Mechanics in Nonequilibrium Thermodynamics]{Stochastic Geometric Mechanics in Nonequilibrium Thermodynamics: Schr\"odinger meets Onsager}

\author{Qiao Huang$^{1,2}$ and Jean-Claude Zambrini$^1$}

\address{$^1$ Group of Mathematical Physics (GFMUL), Department of Mathematics, Faculty of Sciences, University of Lisbon, Campo Grande, Edif\'{\i}cio C6, PT-1749-016 Lisboa, Portugal}
\address{$^2$ Present Address: Division of Mathematical Sciences, School of Physical and Mathematical Sciences, Nanyang Technological University, 21 Nanyang Link, Singapore 637371}
%\ead{qhuang@fc.ul.pt}
\ead{jczambrini@fc.ul.pt}

\vspace{10pt}
\begin{indented}
\item[]\today
\end{indented}

\begin{abstract}
We are describing relations between Schr\"odinger's variational problem and Onsager's approach to nonequilibrium statistical mechanics. Although the second work on reciprocal relations and detailed balance has been published in the same year (1931) as the first one, the impact of Schr\"odinger's idea has not yet been considered in the classical context of Onsager.
\end{abstract}

\vspace{2pc}

\submitto{\JPA}
%\maketitle

\section{Introduction}

The relationship between physical Dynamics and Stochastic Analysis has always been difficult, if not paradoxical. The model of all dynamical theories, in physics, is Classical Mechanics. Its fundamental ordinary differential equations (ODEs) for configuration variables, like Newton one, are second-order in time. When non frictional forces acting on elementary systems are not explicitly time-dependent (i.e., conservative systems) those laws are reversible, or invariant under time reversal.
On the other hand, It\^o's calculus, at the foundations of Stochastic Analysis is essentially a first-order-in-time theory. There are also, of course, second-order stochastic systems, the most famous one being known as Langevin's system. But, in it, the momentum $P$ has to be regarded as a function of the random configuration, (position) $X$, i.e., we need to consider some section of the underlying cotangent bundle (the phase space).

Langevin's system of stochastic differential equations is the prototype of those studied in classical Statistical Mechanics. The stochastic differential equation for $P$ involves, in addition to the classical, conservative, force deriving from a scalar potential, a term proportional to $P$, expressing dissipation of energy, and a Brownian driven fluctuating term. Clearly, Langevin's systems are not invariant under time reversal. In fact, their convergence to an equilibrium (invariant) Gibbs measure is precisely the main source of interest of their study.
Note that, in such a model, typically the interaction between a heavy Brownian particle and a heat bath, the randomness and the dissipation have this interaction as common source (Fluctuation-dissipation Theorem). This is why in the Theoretical (and Mathematical) Physics community, most of researchers tend to identify random fluctuations and dissipation. However, generally, this identification is in no way inevitable. For instance, quantum randomness is often presented as the most irreducible one (to the despair of Einstein). Mathematically, this is due to the crucial role of probability amplitudes in this theory, generally complex functions, which are prerequisites to compute observable (i.e., positive) probabilities of physical events during measurements made by an observer. Since this one is free to choose what measurements should be made, such kind of probability cannot be only an expression of our ignorance of initial conditions, like in classical statistical physics.
In any case, quantum probabilities exist and they are invariant under time reversal, as far as the dynamics of pure states is concerned. Otherwise, any extra randomness of initial quantum states breaks, indeed, the time symmetry invariance of their dynamics.

On the other hand it is also relevant to recall that if the above elementary (pure states) quantum probability exist, it is in the lab and not really in mathematical terms.
The most influential attempt to understand their origin, Feynman's Path Integrals approach, involves, in the simplest case, diffusion processes with complex diffusion coefficients whose inexistence has been shown long ago (although much has been learned about mathematizations of Feynman's ideas since their publication \cite{AHM76}).

The experience of Quantum Field theory has accustomed us to think in terms of an ``Euclidean'' detour (i.e., the Wick-rotated or statistical mechanics one) when looking for quantum probabilistic results. What we are going to do in this work is to describe a geometric and stochastic dynamical theory founded on what was probably the first Euclidean approach to quantum probabilities. Due to Schr\"odinger, in 1931--32, it was almost forgotten until the mid-eighties, when its relations with subsequent attempts to interpret those probabilities (R. Feynman, M. Kac, E. Nelson...) were discovered \cite{Zam86}. In recent years, this circle of ideas spread to a number of scientific communities, in particular, Mass Transportation \cite{Leo14}, Probability \cite{Mik21,LRZ14}, Functional Analysis \cite{VC11}, Geometric Hydrodynamics \cite{KMM21}, and also Economics \cite{Gal18}.

Although the viewpoints are different, there are some similarities between our approach and what is known as Koopman(-von Neumann) method \cite{Koo31,VN32}. There, the main idea was to use tools of operators in Hilbert space for the description of classical dynamical systems.

It is relevant to observe that exactly at the time of Schr\"odinger's observation which is at the origin of our approach (1931-32), Koopman and von Neumann published theirs, partly rediscovered long after by Della Riccia and Wiener \cite{DRW66}. If this last work was not very influential (although motivated by Wiener's early interest in the probabilistic content of quantum mechanics) Koopman's operator method has become fashionable again in various contexts: machine learning, fluid dynamics etc. in addition to its original aim. For a recent application in the study of classical-quantum correlation dynamics, cf. \cite{BGBT19}.

Here we shall use methods of (geometric) stochastic analysis in the approach of some aspects of nonequilibrium classical statistical mechanics suggested by the solution of the problem formulated initially by Schr\"odinger. They will be strongly inspired by the quantization of classical systems, which was the main concern of the father of wave mechanics.

But our approach will also relate to Onsager's famous work on what he called ``reciprocal relations'' in statistical physics and thermodynamics (curiously also published in 1931 \cite{Ons31a,Ons31b}), whose relevance in classical Lagrangian manifold theory has been already noticed long ago \cite[Section 5.3]{AM78}.

In two papers (1931--31) still largely unknown or misunderstood, Schr\"odinger suggested an analogy between the recently discovered Wave Mechanics and classical Statistical Physics. This analogy had been suggested by his reading of a Gifford lecture of A.S. Eddington (1926--27). Schr\"odinger formulated a variational problem whose solution should be classical diffusions with probability densities of a form very reminiscent of Born's interpretation in quantum theory. Convinced that his problem had a solution, Schr\"odinger concluded: ``It was so striking when I found it that it is difficult for me to believe it was purely accidental.''

In the second section of this paper, we shall present an overview of Schr\"odinger's variational problem and its solution in term of a class of time-reversible but generally inhomogeneous diffusions. They satisfy a nonequilibrium generalization of Kolmogorov's traditional reversibility condition (recalled afterwards in Section \ref{sec-5} in the context of Langevin equation). The underlying class of (``Bernstein's reciprocal'') stochastic processes will be defined together with a couple of others, more recent, variational characterizations.

The third section summarizes Onsager's approach of thermodynamic systems slightly out of equilibrium, leading to his famous reciprocal relations. Those have been proven relevant in a number of scientific fields, including Geometric Mechanics.

Section \ref{sec-StochGeom} introduce a Stochastic Geometric Mechanics, with Hamiltonian and Lagrangian formulations, where the above mechanical version of Onsager's reciprocal relations, regarded as a Riemannian structure on thermodynamic space, plays an important role. The key technical tool, in this section, is a ``second-order'' differential geometry, due to Schwartz-Meyer and used here as a kind of quantization method for classical differential geometry.

In Section \ref{sec-5}, we focus on one of the central themes of our paper, the traditional versions of reversibility in classical statistical physics and in quantum mechanics of pure states. We compare them with the nonequilibrium one advocated by Schr\"odinger in 1931, for the needs of his variational problem in statistical physics. We also explain there that a number of results of Section 4 are consistent with another dynamical interpretation of Schr\"odinger's idea, namely, a kind of Euclidean analogy with quantum mechanics for pure states, involving a well defined class of stochastic processes (here, diffusions) whose properties are qualitatively close to the informal ones used by Feynman in his path integrals approach.

The last section mentions some prospects regarding the various notions of entropies underlying the solution of Schr\"odinger's variational problem, whose target was Markovian processes. More generally, Bernstein's reciprocal but non-Markovian processes should be of interests in the study of quantum (Euclidean) nonequilibrium statistical phenomena.

\section{Schr\"odinger's problem}\label{sec-Schrodinger}

%As an important branch of quantum physics, wave mechanics is to use the Schr\"odinger equation to study quantum mechanical systems and make predictions. Schr\"dinger himself was not entirely comfortable with the implications of quantum theory referring to his theory as ``wave mechanics''. He wrote much about the probability interpretation of quantum mechanics. The most famous attempt of him to challenge the Copenhagen interpretation is the thought experiment called Schrödinger's cat paradox.

A still little-known attempt by Schr\"odinger to question some of the foundations of quantum mechanics was published in 1931 and 1932. It was devoted to an analogy between wave mechanics and statistical mechanics. There he used two heat equations, one for forward diffusions and the other for backward, to deduce a formula that is very similar to Born's probabilistic interpretation of Schr\"odinger equation. He said that it was ``so striking to me when I found it, that it is difficult for me to believe it purely accidental.''

\subsection{Schr\"odinger's original idea and some consequences}%\label{sec-2-1}

Instead to repeat what has been done and elaborated in the last 35 years \cite{Zam15}, we summarize briefly here the main results relevant to our present stochastic dynamical purpose.

Schr\"odinger considers the simplest system of classical Brownian particles whose dynamics is described by the self-adjoint Hamiltonian operator
\begin{equation}\label{Hamiltonian}
  \mathcal H = \mathcal H^* = -\frac{1}{2} \Delta + U(x),
\end{equation}
on $L^2(\R^n)$, where $U:\R^n\to\R$ is a bounded below potential.
Observing that the associated quantum unitary group $\exp (-it \mathcal H)$ can be analytically continued, via $t \to -it$, into the self-adjoint semigroup in $L^2(\R^n)$:
\begin{equation*}
  T(t) = e^{-t \mathcal H},
\end{equation*}
so that, for any positive $\psi$ in a dense domain of $\mathcal H$, $w^*_\psi(x,t) = (e^{-(t-t_0) \mathcal H} \psi) (x)$ solves the initial value problem, $t\ge t_0$
\begin{equation}\label{backward-heat}
  -\frac{\pt w^*}{\pt t} = \mathcal H^* w^*, \quad w^*(t_0,\cdot) = \psi.
\end{equation}
Schr\"odinger referred to this first Euclidean step as a banal ``superficial analogy'' with his wave equation. Strangely enough, for decades, his few readers will not go beyond this point.

But Schr\"odinger takes also note of the critical fact that, if $t\le t_1$, we could define as well, on appropriate domain, a final value problem for $w_\phi(x,t) = (e^{-(t_1-t) \mathcal H} \phi) (x)$, namely,
\begin{equation}\label{forward-heat}
  \frac{\pt w}{\pt t} = \mathcal Hw, \quad w(t_1,\cdot) = \phi.
\end{equation}
Let us stress that our $*$ notation does not mean that $w^*$ and $w$ are trivially related as for complex $L^2$ functions. Conditions on $\psi$ and $\phi$ will be given later.

Now, considering the same system, for $t\in I=[t_0,t_1]$, Schr\"odinger's original thought experiment and problem \cite{Sch32} was:
\begin{quote}
Imagine a Brownian particle with Hamiltonian $\mathcal H$ evolving from time $t_0$ with an initial distribution $\rho_{t_0}$. Suppose we observed the state of the particle at time $t_1>t_0$ with another distribution $\rho_{t_1}$. What is the ``most likely'' evolution of the probability distribution $\rho_t$ of this particle at time $t\in [t_0,t_1]$?
\end{quote}

Schr\"odinger used an entropic argument and a discretization technique, familiar in statistical physics, but he knew very well what he wanted to achieve and suggest. His justification has been made rigorous many times since the sixties. As a sample cf. \cite{Beu60,Fol88,CWZ00,Leo14,Mik21}. His result is that this most probable diffusion, say $X(t)$, satisfies a \emph{Born-type formula}, that is, for any Borelian $U$,
\begin{equation}\label{Bonn}
  \P(X(t)\in U) = \int_U w w^* (t,x) dx,
\end{equation}
or, equivalent, the probability distribution $\rho_t$ has a specific (integrable) product form
\begin{equation}\label{Bonn-1}
  \rho_t(dx) = w w^* (t,x) dx,
\end{equation}
where $w^*$ and $w$ are positive solutions of \eref{backward-heat} and \eref{forward-heat} respectively.

First, a remark about the physical interpretation of Schr\"odinger's idea. He is clearly describing relatively rare events. For $\mathcal H$ we expect, on the basis of our macroscopic experience, irreversible dynamics given by \Eref{backward-heat}. But, in fact, any good physics experiment is organized in such a way that mostly interesting events are collected. And without a careful organization of the set-up, they could be pretty rare. In this sense, any physics experiment is very conditioned. This plays a key role in the interpretation of the consequences of Schr\"odinger ``gedanken experiment''.

With the benefits of knowing 90 years of scientific history since Schr\"odinger's observation, we can affirm now what he could not guess, back then. His suggestion was a paradigm shift about stochastic dynamics, whose impact was accidentally delayed for historical reasons.

The unorthodox data of $\rho$ at the boundary $\pt I$ of a time interval has indeed remarkable consequences. Formally, $w$ solving \eref{forward-heat} is a kind of time reversal of $w^*$. Let us see why this is more than a formal relation. Since $\{\rho_{t_0}(dx), \rho_{t_1}(dz)\}$ are arbitrary, this diffusion should not be time-homogeneous in general. It has been proved (e.g. \cite{CWZ00,Zam86}, or \cite{Jam75} for the forward filtration case) that its forward and backward transition probabilities for $s\le t\in I=[t_0,t_1]$ are, respectively,
\begin{eqnarray}
  p(s,x;t,dy) &= h(x, t-s, y) \frac{w(t,y)}{w(s,x)} dy, \label{forward-ts} \\
  p^*(s,dx;t,y) &= \frac{w^*(s,x)}{w^*(t,y)} h(x,t-s,y) dx, \label{backward-ts}
\end{eqnarray}
where $h$ denotes the heat kernel $h(x, t-s, y) = (e^{-(t-s) \mathcal H} \delta_x)(y)$ associated with \Eref{backward-heat}. Given the product form in \eref{Bonn} $\rho(x,t) = ww^* (t,x)$ of Schr\"odinger's most probable process, his nonequilibrium generalization of detailed balance condition (see also \cite{Zam86,Zam15}) becomes
\begin{equation}\label{noneq-balance}
  \rho(x,s) p(s,x,t,dy) dx = p^*(s,dx,t,y) \rho(y,t) dy, \quad s\le t \in I,
\end{equation}
which also means that the joint distribution of the Markovian time-reversible diffusion (in Schr\"odinger sense) of two instants $s\le t\in I$ is of the specific form
\begin{equation}\label{joint-prob}
  \mu_{s,t}(dx,dy) = w^*(s,x) h(x, t-s, y) w(t,y) dxdy
\end{equation}
for $\{w^*(t_0), w(t_1)\} = \{\psi,\phi\}$ unspecified (but positive) boundary conditions of the two heat equations \eref{backward-heat} and \eref{forward-heat}. Since $h$ is positive, these boundary conditions must be as well, for a meaningful joint probability $\mu$.

By construction, the marginals of $\mu$ should therefore satisfy the system for $\{\psi,\phi\}$,
\begin{equation}\label{Sch-system}\left\{
  \eqalign{
    \psi(x) \int h(x,t_1-t_0,y) \phi(y) dy = \rho_{t_0}(x), \\
    \phi(y) \int h(x,t_1-t_0,y) \psi(x) dx = \rho_{t_1}(y),
  }\right.
\end{equation}
as found originally by Schr\"odinger. Also notice that the joint distribution \eref{joint-prob} is the Euclidean version of what Feynman will call, years after, a (complex) ``transition element'' in \cite{FH65}. It is crucial to stress that $\psi, \phi$ have no other relation between them than to solve the system \eref{Sch-system}, so that \Eref{Bonn} represents, in fact, an Euclidean scalar product.

\subsection{Development in probability: Bernstein's reciprocal processes}\label{sec-2-2}

For a fixed Hamiltonian $\mathcal H$, S. Bernstein sketched a method \cite{Ber32} to construct systematically processes like the ones of Schr\"odinger,
%What Schr\"odinger suggested, as observed by S.~Bernstein one year after him \cite{Ber32}, was to construct processes, for a given diffusion process with Hamiltonian $\mathcal H$ (or more generally, a given Markov processes),
from the data of an ``arbitrary'' joint distribution $\mu_{t_0,t_1}(dx,dy)$ at the frontiers of a time interval $[t_0,t_1]$. He suggested that such stochastic processes could be called \emph{reciprocal processes}, as they are defined in the following intrinsically time-reversible fashion:
\begin{equation*}
  \P(X_t\in A | X_u, X_v, X_{s_1}, \cdots, X_{s_n}) = \P(X_t\in A | X_u, X_v),
\end{equation*}
for any $u<t<v\in[t_0,t_1]$, $\{s_1, \cdots, s_n\} \subset [t_0,u]\cup[v,t_1]$ and Borelian $A$.

This strategy was followed by Jamison \cite{Jam74,Jam75} using Beurling's crucial proof of existence and uniqueness of solutions $(\psi,\phi)$ for system \eref{Sch-system} (cf. \cite{Beu60} and Section \ref{sec-5}). %after decades of silence.
He observed that, in general, such processes have no reason to be Markovian. They are Markovian when and only when the end-point joint distribution $\mu_{t_0,t_1}$ has the form \eref{joint-prob} advocated by Schr\"odinger, i.e.,
\begin{equation}\label{Markov-marginal}
  \mu_{t_0,t_1}(dx,dy) = \psi(x) h(x, t_1-t_0, y) \phi(y) dxdy.
\end{equation}
In the latter case, in particular, if the Hamiltonian $\mathcal H$ is of the following diffusion-type, with constant diffusive coefficients $(D^{\alpha\beta})$,
\begin{equation*}%\label{Hamiltonian-0}
  \mathcal H = - D^{\alpha\beta} \frac{\pt^2}{\pt x^\alpha\pt x^\beta},
\end{equation*}
then resulting reciprocal processes are again diffusions on $I$, with infinitesimal (forward) generators,
\begin{equation}\label{Hamiltonian-recip}
  %\tilde{\mathcal H} = -
  D^{\alpha\beta} \left[ \frac{\pt^2}{\pt x^\alpha\pt x^\beta} + \frac{\pt (\log w)}{\pt x^\beta} \frac{\pt}{\pt x^\alpha} \right],
\end{equation}
where $w$ is the solution of \Eref{forward-heat}.

\subsection{Optimal transport perspective}

F{\"o}llmer used large deviation theory to reformulate Schr\"odinger's variational problem as an entropy minimization problem \cite{Fol88}. Notice that, originally, Schr\"odinger applied the Monte Carlo method, that is to repeat independently his thought experiment a large number of times, say, $N$. From a modern viewpoint, this will produce a sequence of i.i.d. processes $X^i(t)$, $i=1,\cdots, N$, with common law $R$ where $R$ is the law of the Brownian particle with Hamiltonian $\mathcal H$ at time interval $[t_0,t_1]$ and initial distribution $\rho_{t_0}$. The empirical law of the system is
\begin{equation*}
  \L^N = \frac{1}{N} \sum_{i=1}^{N} \delta_{X^i}.
\end{equation*}
Schr\"odinger's problem can be translated into finding the following limit of conditional laws:
\begin{equation}\label{Schrodinger-1}
  \lim_{\e\to0^+} \lim_{N\to\infty} \P\left( \L^N\in \cdot \mid \L^N(b)\in B(\rho_{t_1},\e) \right),
\end{equation}
where $B$ means the ball in the space of all probability laws of processes on $[t_0,t_1]$ with initial marginal distribution $\rho_{t_0}$, equipped with some compatible distance.

A straightforward application of \emph{Sanov's theorem} implies that for any ``regular'' subset $A$ of that space,
\begin{equation*}
  \P\left( \L^N\in A \right) \stackrel{n\to\infty}{\asymp} \exp\left[ - n \left( \inf_{P\in A} D_{\mathrm{KL}}(P|R) \right) \right],
\end{equation*}
where $D_{\mathrm{KL}}$ is the \emph{Kullback–Leibler divergence} (also called relative entropy) defined by
\begin{equation*}
  D_{\mathrm{KL}}(P|R) =
    \cases{
      \E_P \left[ \log\left( \frac{dP}{dR} \right) \right], & $P \ll R$, \\
      +\infty, & otherwise. }
\end{equation*}
So that the limit in \Eref{Schrodinger-1} is
\begin{equation*}
  \lim_{\e\to0^+} \lim_{N\to\infty} \P\left( \L^N\in \cdot \mid \L^N(t_1)\in B(\rho_{t_1},\e) \right) = \delta_{P_*},
\end{equation*}
where $P_*$ is the minimizer of $D_{\mathrm{KL}}(P|R)$ over all probability laws $P$ whose two endpoint marginal distributions are given by $\rho_{t_0}$ and $\rho_{t_1}$. The time $t$ marginal of $P_*$ reduces to the solution $\rho_t$ of Schr\"odinger's problem in \Eref{Bonn-1}.

As an analogue of Benamou-Brenier formula in classical Optimal Transport theory, the previous relative entropy minimization problem, with Hamiltonian $\mathcal H$ given by \eref{Hamiltonian} for example, is equivalent to minimize the following average action \cite{Leo14},
\begin{equation}\label{Benamou-Brenier}
  \int_{t_0}^{t_1} \int_{\R^n} \frac{|v(t,x)|^2}{2} \rho(t,dx) dt
\end{equation}
among all pairs $(\rho,v)$ solving the Fokker-Planck equation,
\begin{equation*}%\label{FP-eqn}
\left\{
  \eqalign{
    &\frac{\pt}{\pt t} \rho + \divg \left(\rho v \right) - \frac{1}{2} \Delta \rho = 0, \\
    &\rho(t_0) = \rho_{t_0}, \ \rho(t_1) = \rho_{t_1}.}\right.
\end{equation*}
The minimizer of \eref{Benamou-Brenier} is the pair $(\rho,\nabla S)$ where $S$ solves the following Hamilton-Jacobi-Bellman (HJB) equation,
\begin{equation}\label{HJB}\left\{
  \eqalign{
  & \frac{\pt S}{\pt t} + \frac{1}{2} |\nabla S|^2 + \frac{1}{2} \Delta S = 0, \\
  & S(t_1) = \log\phi.}\right.
\end{equation}
In fact, this HJB equation is related to the heat equation \eref{forward-heat} via the Cole-Hopf transformation (also called logarithmic transformation in stochastic optimal control \cite{FS06}):
\begin{equation}\label{Cole-Hopf}
  S = \log w.
\end{equation}

\section{Onsager's theory}

The theory of fluctuations of Brownian particles, can be generalized to any random quantity, subjected to uncorrelated fluctuations, such as the thermal fluctuations of a thermodynamic system at or close to its equilibrium state. More precisely, as well as the position of a Brownian particle or the velocity of a Langevin particle, a thermodynamic observable $X(t)$ fluctuates in time at thermal equilibrium. Such fluctuations, as it happens for the Brownian particle, are originated by the interactions with an exceedingly large number of microscopic degrees of freedom, present in the system. In practice, one assumes that at thermal equilibrium all these microscopic degrees of freedom evolve as independent stochastic variables, also independent of the actual value of $X(t)$. Due to their large number, relative fluctuations are expected to be quite small and eventually vanish in the thermodynamic limit for any macroscopic observable.

Onsager's Nobel Prize contribution is the discovery of his reciprocal relations, which ``represent a further law making a thermodynamic study of irreversible processes possible''. Readers can refer to \cite{Rei16,Oon17,LP17,Gas22} for more about Onsager's theory.

\subsection{Equilibrium fluctuations}

Suppose that we study a system that requires $n+1$ thermodynamic coordinates (i.e., its thermodynamic space is $n+1$-dimensional), consisting of $n$ work coordinates $(X^i)$ and the internal energy $E=:X^0$. The variables $(X^i)$ are extensive quantities that we can observe and control mechanically (and electromagnetically) in a macroscopic way. Each $X^i$ can be associated to a intensive quantity $Y_i$ conjugate to it (with respect to the internal energy). These constitute a system of $n$ conjugate pairs $(Y_i, X^i)$, $i=1,\cdots, n$. The most commonly considered conjugate pairs are: pressure/volume pair $(-P, V)$, chemical potential/particle number pair $(\mu, N)$, magnetic field/moment pair $(\mathbf B, \mathbf M)$, etc. The temperature/entropy pair $(T, S)$ is not included here, as the entropy $S$ is not considered as a work coordinate.

The first law of thermodynamics is essentially the conservation of energy of the system. The change of the internal energy $E$ cannot be explained solely in terms of the total work $W$ supplied to the system, and the discrepancy $Q$ is understood as the energy transferred in the form of heat to the system:
\begin{equation*}
  dE= \delta Q + W.
\end{equation*}
If the system is closed and isolated, the first law of thermodynamics yields the conservation of internal energy $E$ of the system.

The fundamental thermodynamic relation of Gibbs expresses $E$ as a function of $S$ and work coordinates $(X^i)$. Its differential form is an infinitesimal version of the first law of thermodynamics for quasistatic processes:
\begin{equation*}
  d E=T d S + Y_i dX^i.
\end{equation*}
Here and after, we adopt Einstein's summation convention. The entropic form of Gibbs relation reads
\begin{equation*}
  d S=\frac{1}{T} d E - \frac{Y_i}{T} dX^i.
\end{equation*}
We define the conjugate variables $F_0$ of $E$ and $F_i$ of $X^i$, with respect to entropy, by
\begin{equation*}
  F_0 = \frac{1}{T}, \quad F_i = -\frac{Y_i}{T}.
\end{equation*}
%where $S$ is the entropy, $E$ is internal energy, $T$ is absolute temperature, $P$ is the hydrostatic pressure, $V$ is the volume, $\mu$ is chemical potential, $N$ particle number, $B$ is the external magnetic field and $M$ the magnetic moment.
Then Gibbs relation reduces to
\begin{equation*}%\label{Gibbs}
  d S = F_\alpha d X^\alpha,
\end{equation*}
where we use the Greek alphabet $(\alpha,\beta,\cdots)$ to indicate the summation from $0$ to $n$, to differ from the Latin alphabet $(i,j,\cdots)$ which ranges from $1$ to $n$. From the chain rule it follows that
\begin{equation*}
  F_\alpha = \frac{\pt S}{\pt X^\alpha}.
\end{equation*}

The entropy $S$ will be a maximum when the system is in an equilibrium state $\bar X = (\bar X^\alpha)$. Any fluctuation around the equilibrium state must cause the entropy
to decrease. We let $(\hat X^\alpha)$ denote the fluctuations
\begin{equation*}
  \hat X^\alpha = X^\alpha - \bar X^\alpha.
\end{equation*}
Assuming the system is close to its equilibrium, the fluctuations are small and we can expand the entropy around its equilibrium value to obtain
\begin{equation}\label{ent}
  S(X) \approx S(\bar X) + \frac{1}{2} A_{\alpha\beta} \hat X^\alpha \hat X^\beta.
\end{equation}
where
\begin{equation*}
  A_{\alpha\beta} = \frac{\pt^2 S}{\pt X^\alpha \pt X^\beta}\bigg|_{X= \bar X} = \frac{\pt^2 S}{\pt \hat X^\alpha \pt \hat X^\beta}\bigg|_{\hat X = 0}.
\end{equation*}
The matrix $(A_{\alpha\beta})$ is symmetric due to Maxwell relations, and it is negative semi-definite since the entropy is a concave function of all the thermodynamic coordinates. \Eref{ent} contains no first-order terms in $X$ to ensure that spontaneous fluctuations about the equilibrium do not cause an increase in the entropy.

%The change in entropy which results from these fluctuations about the equilibrium state is
%\begin{equation*}
%  \delta S := \frac{1}{2} A_{\alpha\beta} \hat X^\alpha \hat X^\beta.
%\end{equation*}
From chain rule it follows that
\begin{equation}\label{force}
  F_\alpha = \frac{\pt S}{\pt X^\alpha} = \frac{\pt S}{\pt \hat X^\alpha} = A_{\alpha\beta} \hat X^\beta.
\end{equation}
We call $(F_\alpha)$ the conjugate variables of the fluctuation variables $(\hat X^\alpha)$.

The generalized flux $J$ is defined as the time derivative of conditional average
\begin{equation*}
  J^\alpha:= \frac{d \langle \hat X^\alpha \rangle_0}{dt},
\end{equation*}
where $\langle \cdot \rangle_0$ stands for the average given the initial value $\hat X^\alpha(0)$.
Then the rate of entropy production due to fluctuations is
\begin{equation}\label{ent-prod-0}
  \frac{d\langle S \rangle_0}{dt} \stackrel{\times}{=} \left\langle \frac{\pt S}{\pt \hat X^\alpha} \right\rangle_0 \frac{d \langle \hat X^\alpha \rangle_0}{dt} = \langle F_\alpha \rangle_0 J^\alpha.
\end{equation}

\begin{remark}\label{remark-1}
  From a modern point of view, Onsager's original derivations are partially incorrect, as indicated by the symbol $\times$ in \Eref{ent-prod-0}. Indeed, stochastic calculus was still unknown in Onsager's time. Since $\hat X$ describe fluctuations, it is a stochastic process in general, modelled by a stochastic differential equation (SDE), see Subsection \ref{sec-3-3}. So in the calculus involving $\hat X$, It\^o's formula must be applied. For example, in the derivation \eref{ent-prod-0} of entropy production rate, we need first apply It\^o' formula to get the stochastic differential of $S$,
  \begin{equation*}
    d S = \frac{\pt S}{\pt \hat X^\alpha} d\hat X^\alpha + \frac{1}{2}\frac{\pt^2 S}{\pt \hat X^\alpha\pt \hat X^\beta} d\hat X^\alpha \cdot d\hat X^\beta,
  \end{equation*}
  where the term $\frac{1}{2}\frac{\pt^2 S}{\pt \hat X^\alpha\pt \hat X^\beta} d\hat X^\alpha \cdot d\hat X^\beta$, additional to \eref{ent-prod-0}, is called It\^o's correction.
%  Then we take expectation and get
%  \begin{equation*}%\label{eqn-5}
%    \frac{d\langle S \rangle_0}{dt} = \frac{\pt S}{\pt \hat X^\alpha} \frac{d \langle \hat X^\alpha \rangle_0}{dt} + \frac{1}{2}\frac{\pt^2 S}{\pt \hat X^\alpha\pt \hat X^\beta} \frac{\langle d\hat X^\alpha \cdot d\hat X^\beta \rangle_0}{dt}.
%  \end{equation*}
\end{remark}

\emph{Boltzmann's entropy formula} provides
\begin{equation}\label{Boltzmann}
  S(X) = k_B \log \Omega(X),
\end{equation}
where $\Omega(X)$ is the number of microstates compatible with the macrostate $X=(X^\alpha)$ in equilibrium. It is clear that the probability for $X$ to have a value in the volume element $dX$ of the thermodynamical space is proportional the number function $\Omega(X)$, i.e., $P(X) \propto \Omega (X) = e^{S(X)/K_B}$. Indeed, we can now substitute \eref{Boltzmann} into \eref{ent} and obtain the following expression for the probability distribution of fluctuations around an equilibrium state,
\begin{equation}\label{ivrt-meas}
  \eqalign{
    P(\hat X) &= \frac{\Omega(X)}{\int\Omega(X) dX} = \frac{1}{\int e^{S(X)/k_B} dX} \exp\left( \frac{S(X)}{k_B} \right) \\
    &= \sqrt{\frac{\det(A)}{(2\pi k_B)^n}} \exp\left( \frac{1}{2k_B} A_{\alpha\beta} \hat X^\alpha \hat X^\beta \right).
  }
\end{equation}
By Laplace transform, one finds the covariance of $\hat X$ as follows,
\begin{equation}\label{cov}
  \langle \hat X^\alpha \hat X^\beta \rangle = -k_B A^{\alpha\beta},
\end{equation}
where $(A^{\alpha\beta})$ is the inverse matrix of $(A_{\alpha\beta})$.

\subsection{Onsager's regression hypothesis and reciprocal relations}

In Onsager's seminal papers \cite{Ons31a,Ons31b}, he assumed that,
\begin{quote}
``... the average regression of fluctuations will obey the same laws as the corresponding macroscopic irreversible process.''
\end{quote}
Nowadays this is known as \emph{Onsager's regression hypothesis}. It means that if the system is in a state fairly close to its equilibrium, i.e., if $\hat X$ may be regarded as small, the conditional average fluctuation $\langle \hat X \rangle_0$ may be expanded as the linear sums
\begin{equation}\label{fluc-exp}
  \frac{d\langle \hat X^\alpha \rangle_0}{dt} \approx -\lambda_\beta^\alpha \langle \hat X^\beta \rangle_0,
\end{equation}
where $\lambda = (\lambda_\beta^\alpha)$ is an $(n+1)\times(n+1)$ constant matrix. Its solution has the following short-time approximation:
\begin{equation}\label{sol-Langevin}
  \langle \hat X^\alpha(t) \rangle_0 = \hat X^\alpha(0) - t \lambda_\beta^\alpha \hat X^\beta(0) + \Or(t^2).
\end{equation}

The principle of \emph{microscopic reversibility} states that the microscopic dynamics of particles and fields is time-reversible, like Newton's equation. According to Boltzmann, this microscopic reversibility implies a principle of detailed balance for collisions. As a consequence, the correlation functions for macroscopic fluctuations $\hat X$ obey the relations
\begin{equation}\label{corr}
  \langle \hat X^\alpha(0) \hat X^\beta(t) \rangle = \langle \hat X^\alpha(t) \hat X^\beta(0) \rangle.
\end{equation}

Substitute \eref{sol-Langevin} into \eref{corr}, and notice that $\langle \hat X^\alpha(0) \hat X^\beta(t) \rangle = \langle \hat X^\alpha(0) \langle \hat X^\beta(t) \rangle_0 \rangle$. We obtain
\begin{equation*}
  \lambda_\gamma^\beta \langle \hat X^\alpha(0) \hat X^\gamma(0) \rangle = \lambda_\gamma^\alpha \langle \hat X^\gamma(0) \hat X^\beta(0) \rangle.
\end{equation*}
Then we apply \Eref{cov} to get
\begin{equation*}
  \lambda_\gamma^\beta A^{\alpha\gamma} = \lambda_\gamma^\alpha A^{\beta\gamma}.
\end{equation*}
Now we can define a new matrix $L^{\alpha\beta} = -\lambda_\gamma^\beta A^{\alpha\gamma}$ and obtain the following \emph{Onsager's reciprocal relations}:
\begin{equation*}
  L^{\alpha\beta} = L^{\beta\alpha}.
\end{equation*}
The matrix $(L^{\alpha \beta})$ is called the \emph{Onsager matrix}.

If we make use of the generalized force $F$ in \Eref{force}, the time rate
of change of the fluctuation can be written
\begin{equation}\label{eqn-4}
  J^\alpha = \frac{d\langle \hat X^\alpha \rangle_0}{dt} = -\lambda_\beta^\alpha \langle \hat X^\beta \rangle_0 = L^{\alpha\gamma} A_{\gamma\beta} \langle \hat X^\beta \rangle_0 = L^{\alpha\gamma} \langle F_\gamma \rangle_0 = L^{\alpha\gamma} \left\langle \frac{\pt S}{\pt \hat X^\gamma} \right\rangle_0.
\end{equation}
Substituting \eref{eqn-4} in \eref{ent-prod-0}, we obtain the following expression for the entropy production,
\begin{equation}\label{ent-prod-1}
  \frac{d\langle S \rangle_0}{dt} \stackrel{\times}{=} \langle F_\alpha \rangle_0 J^\alpha = L^{\alpha\beta} \langle F_\alpha \rangle_0 \langle F_\beta \rangle_0.
\end{equation}
In out-of-equilibrium conditions the right-hand side of this equation has to be a positive quantity, which vanishes at equilibrium. So the Onsager matrix $(L^{\alpha \beta})$ is positive definite.

\subsection{Generalized fluctuation-dissipation relation}\label{sec-3-3}

In view of \Eref{eqn-4}, let us write the following Langevin-type equation
\begin{equation}\label{generalized-Langevin}
  \frac{d \hat X^\alpha}{dt} = L^{\alpha\beta} \frac{\pt S}{\pt \hat X^\beta} + \xi,
\end{equation}
where $\xi = (\xi^\alpha)$ is a white noise with correlation function
\begin{equation}\label{corr-noise}
  \langle\xi^\alpha(t) \xi^\beta(s)\rangle = 2 D^{\alpha\beta} \delta(t-s),
\end{equation}
and $D=(D^{\alpha\beta})$ is a constant matrix describing the magnitude of the fluctuation.

We know, since Einstein's work on Brownian motion, that $D$ cannot be chosen arbitrarily for a given system, if it is in equilibrium. The noise amplitude $D$ and the dissipation rate $L$ are related.

The (time-independent) Fokker-Planck equation for \Eref{generalized-Langevin} in equilibrium reads (the drift is $L\nabla S$ and the diffusion matrix is $2D$),
\begin{equation}\label{FK}
  D^{\alpha\beta} \frac{\pt^2 \rho}{\pt x^\alpha \pt x^\beta} - L^{\alpha\beta} \frac{\pt}{\pt x^\alpha} \left( \rho \frac{\pt S}{\pt x^\beta} \right) = 0.
\end{equation}
Since \eref{ivrt-meas} is a solution of \Eref{FK}, we deduce
\begin{equation}\label{fluc-diss}
  D^{\alpha\beta} = k_B L^{\alpha\beta}.
\end{equation}
This relation is a generalization of the \emph{fluctuation-dissipation relation} in classical Brownian motion theory.

\begin{remark}\label{remark-2}
  For the same reason as Remark \ref{remark-1}, the calculation of \Eref{ent-prod-1} is not completely correct.
%  By \eref{generalized-Langevin}, \eref{corr-noise} and \eref{fluc-diss}, we know that $d\hat X^\alpha(t) \cdot d\hat X^\beta(t) = 2 k_B L^{\alpha\beta} t$. Then \eref{ent-prod-1} should reads, using \eref{eqn-5},
%  \begin{equation*}
%    \frac{d\langle S \rangle_0}{dt} = L^{\alpha\beta} \left( \langle F_\alpha \rangle_0 \langle F_\beta \rangle_0 + k_B \frac{\pt^2 S}{\pt \hat X^\alpha\pt \hat X^\beta} \right).
%  \end{equation*}
  On the other hand, the close-to-equilibrium approximations used in \eref{ent} and \eref{fluc-exp} are important to achieve \Eref{generalized-Langevin}. All of these show that Onsager's original derivations are a bit defective. What we are going to do in the upcoming section \ref{sec-StochGeom} is to reformulate Onsager's idea from Schr\"odinger's viewpoint and fix those problems.
\end{remark}

From a modern point of view, what Onsager is telling us is that mesoscopic observables obey the large deviation principle around the macroscopic law due to the law of large numbers:
\begin{equation*}
  \P\left[ \frac{1}{\delta t} \int_0^{\delta t} \left(\frac{d X(s)}{d s}- J(t) \right) ds \in dx \right] \approx 4d \exp\left(- \frac{x^T D^{-1} x}{4d} \delta t \right),
\end{equation*}
where $d$ is the spatial dimension.

\subsection{From Schr\"odinger to Onsager}

From \eref{generalized-Langevin}, \eref{corr-noise} and \eref{fluc-diss}, as well as Boltzmann's formula \eref{Boltzmann}, we know that the infinitesimal generator of the fluctuation process $\hat X$ is
\begin{equation*}
  D^{\alpha\beta} \left[ \frac{\pt^2}{\pt x^\alpha\pt x^\beta} + \frac{\pt (\log\Omega)}{\pt x^\beta} \frac{\pt}{\pt x^\alpha} \right],
\end{equation*}
where $\Omega$ is the number function in \Eref{Boltzmann}. This generator is exactly of the form \eref{Hamiltonian-recip}. This suggests that Onsager's way to approach nonequilibrium statistical physics can be compatible with Schr\"odinger's thought experiment.

Indeed, we can replace the Brownian particles of Schr\"odinger's problem by a thermodynamic system close to its equilibrium state, and the position variables of the particle by the deviation $(\hat X^\alpha)$ of thermodynamic coordinates from the equilibrium. From time $t_0$, the observables $(\hat X^\alpha)$ start to spontaneously evolve, with an initial distribution near to the equilibrium distribution \eref{ivrt-meas}. At a sufficiently large time $t_1$, the system reaches the equilibrium state and the observables $(\hat X^\alpha)$ end up with the equilibrium distribution \eref{ivrt-meas}.

Since the time scale of $t_1$ is very large, mathematically we may treat it as infinity. The heat equation \eref{forward-heat} and its associated HJB equation \eref{HJB} becomes approximately time-independent, so the Cole-Hopf transformation \eref{Cole-Hopf} has the same form as Boltzmann's entropy formula \eref{Boltzmann}. This suggests that the entropy $S$, as a function of fluctuation variables $(\hat X^\alpha)$, should satisfy the HJB equation. %and the solution of heat equation \eref{forward-heat} are related to the number function via $w=\Omega^{k_B}$.

Moreover, the fact that the Onsager matrix $(L^{\alpha \beta})$ is symmetric and positive definite suggests that one can treat $(L^{\alpha \beta})$ as a Riemannian metric on the thermodynamic space of fluctuation variables $(\hat X^\alpha)$. The conjugation between $(\hat X^\alpha)$ and $(F_\alpha)$ and their relation with $S$ in \eref{force} indicate that Legendre transform should underlie the framework.

In view of these observations and Remarks \ref{remark-1} and \ref{remark-2}, it is possible to connect Schr\"odinger's problem with Onsager's theory in systematical way, compatible with It\^o's stochastic calculus. In our paper \cite{HZ22}, we have developed a stochastic framework of geometric mechanics, based on the theory of second-order differential geometry due to L. Schwartz and P.-A. Meyer \cite{Eme07,Mey81}. These geometric theories are indeed motivated by and compatible with the stochastic nature of Brownian randomness and It\^o's calculus.

\section{Stochastic geometric mechanics: beyond Schr\"odinger and Onsager}\label{sec-StochGeom}

In this section, we review our framework of stochastic geometric mechanics, of which details can be found in \cite{HZ22}. One can also find a brief review of this framework in \cite{HZ22a}. Then we apply this framework to Schr\"odinger's problem with Onsager's motivation.

\subsection{Second-order differential geometry}

Consider a general $m$-dimensional manifold $M$, which can be the flat Euclidean spaces, or the thermodynamic space. The coordinates on $M$ are denoted by $(x^\alpha)$.

The second-order (SO) tangent bundle $\mathcal T^S M$ is a fiber bundle such that its sections are second-order operators of the following form
\begin{equation}\label{so-tangent}
  A = A^\alpha \frac{\partial}{\partial x^\alpha} + A^{\alpha\beta} \frac{\partial^2}{\partial x^\alpha\partial x^\beta}.
\end{equation}
A natural frame of it is the following set of elementary differential operators of first-order and second-order:
\begin{equation}\label{frame}
  \left\{ \vf{x^\alpha}, \frac{\pt^2}{\pt x^\alpha \pt x^\beta}: 1\le \alpha\le \beta \le m \right\}.
\end{equation}
The dual bundle is the second-order cotangent bundle, denoted by $\mathcal T^{S*} M$, of which the frame dual to \eref{frame} is
\begin{equation}\label{co-frame}
  \left\{ d^2 x^\alpha, \textstyle{\frac{1}{2}} dx^\alpha\cdot dx^\alpha, dx^\alpha\cdot dx^\beta: 1\le \alpha< \beta \le m \right\},
\end{equation}
where $d$ is the classical differential operator, $d^2$ is called second-order differential operator and $\cdot$ is called symmetric product. They acts on smooth functions $f, g$ on $M$ as follows,
\begin{equation*}
  \langle d^2 f, A \rangle = \frac{\pt f}{\pt x^\alpha} A^\alpha + \frac{\pt^2 f}{\pt x^\alpha \pt x^\beta} A^{\alpha\beta}, \quad \langle df\cdot dg, A \rangle = \frac{\pt f}{\pt x^\alpha} \frac{\pt g}{\pt x^\beta} A^{\alpha\beta},
\end{equation*}
where $A$ is the operator \eref{so-tangent}. From \Eref{co-frame} and after, we adopt the convention that $dx^\beta\cdot dx^\alpha = dx^\alpha\cdot dx^\beta$ for all $1\le \alpha< \beta \le m$. The factor $\frac{1}{2}$ in front of in $dx^\alpha\cdot dx^\alpha$ in \eref{co-frame} is a normalized constant since the off-diagonal elements $dx^\alpha\cdot dx^\beta$ are folded to the lower triangular ones with $\alpha< \beta$.
The coordinates $(x^\alpha)$ on $M$ induce a canonical coordinate system on $\mathcal T^{S*} M$, denoted by $(x^\alpha, p_\alpha, o_{\alpha\beta})$, where extra variables $(o_{\alpha\beta})$ describe second-order effects. Sections of $\mathcal T^{S*} M$ are called second-order forms.

As indicated by the notation, the expression $d^2 f$ can indeed be understood as $d(d f)$, where the differential $d$ inside the parentheses is de Rham's exterior differential on $M$, while the one outside is the exterior differential on $TM$ by regarding the first differential $df$ as a function on $TM$ \cite{LT99}. The bundle $\mathcal T^{S*} M$ is not an unacquainted object, since $\mathcal T^{S*} M\times\R$ is bundle diffeomorphic to the classical second-order jet bundle of the trivial bundle $(M\times \R, \pi, M)$. The bundle $\mathcal T^S M$ can also be endowed with a jet-like structure by introducing the notion of `stochastic jets' for diffusion processes, as opposed to the higher-order tangent bundle of higher-order jets for deterministic curves introduced in \cite{GHR11}. The reason why we make use of the two odd-looking bundles is that, according to Schwartz and Meyer's heuristic principle, any geometric statement for such second-order (co)tangent vectors will have a probabilistic content. See \cite{HZ22} for more details.

Let be given a probability space $(\Omega,\mathcal F,\P)$ and a nondecreasing filtration $\Pred_t\subset \mathcal F$. Roughly speaking, $\mathcal F$ contains all the probabilistic information while $\Pred_t$ only contains all past information before the present time $t$. For a diffusion process $(x^\alpha(t))$ valued on $M$, the coefficients of its generator can be characterized by
\begin{equation}\label{mean-d-x}
    (D^\alpha x)(t) := \lim_{\e\to0^+} \E \ts{\left[ \frac{x^\alpha(t+\e)-x^\alpha(t) }{\e} \Big| \Pred_t \right]},
\end{equation}
\begin{equation*}
    (Q^{\alpha\beta} x)(t) := \lim_{\e\to0^+} \E \ts{\left[ \frac{(x^\alpha(t+\e)-x^\alpha(t)) (x^\beta(t+\e)-x^\beta(t))}{\e} \Big| \Pred_t \right]}.
\end{equation*}
The pair $(D^\alpha x(t),Q^{\alpha\beta} x(t))$ is a process taking values in $\mathcal T^S M$, and called the \emph{mean derivatives} of $(x^\alpha(t))$. The mean derivative of a function $f$ acting on $x(t)$ can be defined similarly and expressed by
\begin{equation}\label{mean-d}
  \D_t f := \lim_{\e\to0^+} \E \ts{\left[ \frac{f(x(t+\e))-f(x(t)) }{\e} \Big| \Pred_t \right]} = \frac{\pt f}{\pt t} + D^i x \frac{\pt f}{\pt x^i} + \frac{1}{2} Q^{jk} x \frac{\pt^2 f}{\pt x^j \pt x^k}.
\end{equation}

In general, $(D^\alpha x)$ does not transform as a vector field, which can be verified by applying It\^o's formula for changes of coordinates. In order to overcome this problem, we equip $M$ with a linear connection $\nabla$, and use it to compensate the correction term resulting from It\^o's formula. That is, we define the following $\nabla$-dependent mean derivative, in forms of Christoffel's symbols,
\begin{equation*}%\label{mean-d}
  D_\nabla^\alpha x = D^\alpha x + \ts{\frac{1}{2}} \Gamma^\alpha_{\beta\gamma} Q^{\beta\gamma}x.
\end{equation*}
Then $D_\nabla X$ does transform as a vector field.

\subsection{Stochastic Hamiltonian mechanics}

The SO Poincar\'e form $\theta$ is a SO form on $\mathcal T^{S*} M$, given by
\begin{equation*}
  \theta = p_\alpha d^2 x^\alpha + \textstyle{\frac{1}{2}} o_{\alpha\beta} dx^\alpha\cdot dx^\beta.
\end{equation*}
And the canonical SO symplectic form becomes
\begin{equation*}
  \omega = d^2 \theta = d^2 x^\alpha \wedge d^2 p_\alpha + \textstyle{\frac{1}{2}} dx^\alpha\cdot dx^\beta \wedge d^2 o_{\alpha\beta}.
\end{equation*}
See \cite[Section 6.2]{HZ22} for the definition of the action of the above SO form $\omega$ on SO operators of the form \eref{so-tangent}. It can be verified that $\omega$ is nondegenerate in the sense that $\omega(A,B)=0$ for all SO operators $B$ implies $A=0$. We call the pair $(\mathcal T^{S*} M, \omega)$ a second-order symplectic manifold. A SO Hamiltonian is a smooth function $H: \mathcal T^{S*} M\times\R \to \R$. Its SO Hamiltonian vector field is a second-order vector field on $\mathcal T^{S*} M$ given by
\begin{equation*}
  A_H \approx \frac{\pt H}{\pt p_\alpha} \vf{x^\alpha} - \frac{\pt H}{\pt x^\alpha} \vf{p_\alpha} + \frac{\pt H}{\pt o_{\alpha\beta}} \frac{\pt^2}{\pt x^\alpha \pt x^\beta} - \frac{\pt^2 H}{\pt x^\alpha \pt x^\beta} \vf{o_{\alpha\beta}},
\end{equation*}
where $\approx$ stands for principal parts. The stochastic Hamilton's equations are the following SDEs on $\mathcal T^{S*} M$,
\begin{equation}\label{stoch-Hamilton-eqns}
  \left\{
  \eqalign{
    D^\alpha x &= \frac{\pt H}{\pt p_\alpha}, \\
    Q^{\alpha\beta} x &= 2\frac{\pt H}{\pt o_{\alpha\beta}}, \\
    D_\alpha p &= - \frac{\pt H}{\pt x^\alpha}, \\
    D_{\alpha\beta} o &\approx - \frac{\pt^2 H}{\pt x^\alpha \pt x^\beta}, \\
%    C_{ij} \frac{\pt H}{\pt o_{ij}} &= \frac{1}{2} Q_{jk} p \frac{\pt^2 H}{\pt p_j \pt p_k} + \frac{1}{2} Q_{ijkl} o \frac{\pt^2 H}{\pt o_{ij} \pt o_{kl}} + Q^j_k(x,p) \frac{\pt^2 H}{\pt x^j \pt p_k} \\
%    &\quad + Q^j_{kl}(x,o) \frac{\pt^2 H}{\pt x^j \pt o_{kl}} + Q_{jkl}(p,o) \frac{\pt^2 H}{\pt p_j \pt o_{kl}},
  }\right.
\end{equation}

The above equations are not necessarily solvable. But if we set $p_\alpha=p_\alpha(t,x)$ and $o_{\alpha\beta}=o_{\alpha\beta}(t,x)$, they can be simplified. Indeed, the first two equations give the generator of $(x^\alpha(t))$, which can be used to apply It\^o's formula to the last two equations. It follows that
\begin{equation*}%\label{Maxwell}
  o_{\alpha\beta} = \frac{\pt p_\alpha}{\pt x^\beta} = \frac{\pt p_\beta}{\pt x^\alpha}.
\end{equation*}
These are known as \emph{Maxwell's relations} in thermodynamics (also called  Onsager's relations somewhat inaccurately in \cite{AM78}).

A change of coordinates leaving the form of the stochastic Hamilton's equations \eref{stoch-Hamilton-eqns} unchanged give rise to the following SO Hamilton-Jacobi-Bellman equation,
\begin{equation}\label{HJB-0}
  \frac{\pt S}{\pt t} + H(d^2 S,t) = \frac{\pt S}{\pt t} + H\left( x^\alpha, \frac{\pt S}{\pt x^\alpha}, \frac{\pt^2 S}{\pt x^\alpha \pt x^\beta}, t \right) = 0.
\end{equation}
This HJB equation can also be related to the stochastic Hamilton's equations \eref{stoch-Hamilton-eqns} via the following statement: for any process $x(t)$ satisfying the first two equations of \eref{stoch-Hamilton-eqns}, the process
\begin{equation*}
  (x^\alpha(t), p_\alpha(t), o_{\alpha\beta}(t) ) = \left( x^\alpha(t), \frac{\pt S}{\pt x^\alpha}(t, x(t)), \frac{\pt^2 S}{\pt x^\alpha\pt x^\beta}(t, x(t)) \right)
\end{equation*}
solves \eref{stoch-Hamilton-eqns} if and only if $S$ solves \eref{HJB-0}.

There is no unique way to obtain a second-order Hamiltonian from a given classical Hamiltonian $H_0: T^* M\times\R \to \R$. But if we specify a Riemannian metric $g$, then there is a canonical way, that is,
\begin{equation}\label{canonical-H}
  H(x,p,o,t) = H_0(x,p,t) + \textstyle{\frac{1}{2}} g^{\alpha\beta}(x) \left(o_{\alpha\beta} - \Gamma^\gamma_{\alpha\beta}(x)p_\gamma \right).
\end{equation}
In this case, the stochastic Hamilton's equations can be rewritten to the following global form on $T^* M$,
  \begin{equation}\label{canonical-stoch-Hamilton-eqns}\left\{
    \eqalign{
      D^{}_\nabla x = \nabla_p H_0, \\
      \frac{\overline\D}{dt} p = -d_x H_0,
    }\right.
  \end{equation}
subject to $Q^{\alpha\beta}x(t) = g^{\alpha\beta}(x(t))$, where $\frac{\overline\D}{dt} = \frac{\pt}{\pt t} + \nabla_{D_\nabla X} + \frac{1}{2} \Delta_{\mathrm{LD}}$. The corresponding Hamilton-Jacobi-Bellman equation for this case is
  \begin{equation}\label{HJB-Riem}
    \frac{\pt S}{\pt t} + H_0(d S, t) + \frac{1}{2}\Delta S = 0,
  \end{equation}
where the Laplacian term express clearly the stochastic deformation of classical Hamilton-Jacobi equation.

\subsection{Stochastic Lagrangian mechanics}

The stochastic Lagrangian mechanics is established by considering the following stochastic variational problem: to minimize the action functional
\begin{equation}\label{action}
  \mathcal S[x] = \E \int_{t_0}^{t_1} L_0\left(t, x(t), D_\nabla^{} x(t) \right) dt
\end{equation}
over all diffusion processes $(x(t):t_0\le t\le t_1)$, with initial and final marginal distributions $\mu_0$ and $\mu_1$, and satisfying $Q^{\alpha\beta}x(t) = g^{\alpha\beta}(x(t))$.
%The variation of a diffusion $x(t)$ we used is of the Cameron-Martin type. That is, for an absolutely continuous curves $v:[t_0,t_1]\to T_{x(0)} M$ such that $\int_{t_0}^{t_1} |\dot v(t)|^2 dt < \infty$ and $v(t_0) = v(t_1) = 0$, we define a one-parameter family of diffusions $\{X_\e\}_{\e\in(-\varepsilon,\varepsilon)}$ by the following ODE
%  \begin{equation*}
%    \frac{\pt}{\pt\e}x_\e(t) = \Gamma(x_\e)_0^t v(t), \quad x_0(t) = x(t).
%  \end{equation*}
Solving the stochastic variational problem via the Cameron-Martin type variation leads to the following stochastic Euler-Lagrange equation
\begin{equation}\label{EL}
  \frac{\overline\D}{dt} \left( d_{\dot x} L_0 \right) = d_x L_0.
\end{equation}

It turns out that the stochastic Euler-Lagrange equation is equivalent, as it should, to the global version of stochastic Hamilton's equations \eref{canonical-stoch-Hamilton-eqns} via the Legendre transform
\begin{equation}\label{Legendre}
  D_\nabla^\alpha x = \frac{\pt H_0}{\pt p_\alpha}, \quad L_0(t,x,D_\nabla^{} x) = p_\alpha D_\nabla^\alpha x - H_0(x,p,t).
\end{equation}

For the canonical SO Hamiltonian induced by $H_0$, as in \eref{canonical-H}, we have
\begin{equation*}
  H(x,p,o,t) = p_\alpha D^\alpha x + \frac{1}{2} o_{\alpha\beta} Q^{\alpha\beta} x - L_0(t,x,D_\nabla^{} x).
\end{equation*}
Then we apply the mean derivative \eref{mean-d} to the solution $S$ of the special HJB equation \eref{HJB-Riem},
\begin{equation}\label{DS}
  \D_t S = \frac{\pt S}{\pt t} + D^\alpha x \frac{\pt S}{\pt x^\alpha} + \frac{1}{2} Q^{\alpha\beta} x \frac{\pt^2 S}{\pt x^\alpha \pt x^\beta} =  - H + p_\alpha D^\alpha x + \frac{1}{2} o_{\alpha\beta} Q^{\alpha\beta} x = L_0.
\end{equation}
Hence, the action functional \eref{action} can be expressed as
\begin{equation}\label{action-1}
  \mathcal S[x] = \E \left[ S(t_1,x(t_1)) - S(t_0,x(t_0)) \right].
\end{equation}

\subsection{Back to Schr\"odinger \& Onsager}

Now we take $M$ to be the thermodynamic space of fluctuation variables, and equip it with a Riemannian metric $(L^{\alpha\beta})$. Consider the following SO Hamiltonian
\begin{equation}\label{exmp-Hamiltonian}
  H(x,p,o) = \frac{1}{2} L^{\alpha\beta}(x) p_\alpha p_\beta + k_B L^{\alpha\beta}(x) \left( o_{\alpha\beta} - \Gamma_{\alpha\beta}^\gamma(x) p_\gamma \right),
\end{equation}
which is canonically induced by $H_0 = \frac{1}{2} L^{\alpha\beta}(x) p_\alpha p_\beta$.
%Applying \eref{stoch-Hamilton-eqns} and \eref{Maxwell}, we get
%\begin{equation}
%  \left\{
%  \eqalign{
%    D^\alpha x &= L^{\alpha\beta} p_\beta - \frac{k_B}{2} L^{\beta\gamma} \Gamma_{\beta\gamma}^\alpha, \\
%    Q^{\alpha\beta} x &= k_B L^{\alpha\beta}, \\
%    \left( \frac{\pt}{\pt t} + D^\alpha x \frac{\pt}{\pt x^\beta} + \frac{1}{2} Q^{\beta\gamma} x \frac{\pt^2}{\pt x^\beta\pt x^\gamma} \right) p_\alpha &= - \frac{\pt L^{\beta\gamma}}{\pt x^\alpha} \left[ \frac{1}{2} p_\beta p_\gamma + \frac{k_B}{2} \left( o_{\beta\gamma} - \Gamma_{\beta\gamma}^\delta p_\delta \right)\right] + \frac{k_B}{2} L^{\beta\gamma} \frac{\pt\Gamma_{\beta\gamma}^\delta}{\pt x^\alpha} p_\delta, \\
%    o_{\alpha\beta} &= \frac{\pt p_\alpha}{\pt x^\beta}. \\
%%    C_{ij} \frac{\pt H}{\pt o_{ij}} &= \frac{1}{2} Q_{jk} p \frac{\pt^2 H}{\pt p_j \pt p_k} + \frac{1}{2} Q_{ijkl} o \frac{\pt^2 H}{\pt o_{ij} \pt o_{kl}} + Q^j_k(x,p) \frac{\pt^2 H}{\pt x^j \pt p_k} \\
%%    &\quad + Q^j_{kl}(x,o) \frac{\pt^2 H}{\pt x^j \pt o_{kl}} + Q_{jkl}(p,o) \frac{\pt^2 H}{\pt p_j \pt o_{kl}},
%  }\right.
%\end{equation}
We first solve the last two partial differential equations of \eref{stoch-Hamilton-eqns} by considering the following backward heat equation:
\begin{equation*}
  \frac{\pt w}{\pt t} + k_B \Delta w = 0,
\end{equation*}
where the Laplacian $\Delta$ is under the Riemannian metric $(L^{\alpha\beta})$. Then the Cole-Hopf transformation
\begin{equation}\label{generating-func}
  S(t,x) = 2k_B \log w(t,x)
\end{equation}
solves the following HJB equation
\begin{equation*}
  \frac{\pt S}{\pt t} + \frac{1}{2} |\nabla S|^2 + k_B \Delta S = 0,
\end{equation*}
where the gradient $\nabla$ is also with respect to $(L^{\alpha\beta})$.
It is easy to verify that
\begin{equation}\label{momenta}
  p_\alpha = \frac{\pt S}{\pt x^\alpha}, \quad o_{\alpha\beta} = \frac{\pt^2 S}{\pt x^\alpha\pt x^\beta}
\end{equation}
solve the last two equations. Therefore, the first two mean differential equations reduce to
\begin{equation}\label{SDE}
  \frac{d x^\alpha}{dt} = L^{\alpha\beta} \frac{\pt S}{\pt x^\beta} + \xi,
\end{equation}
where the white noise $\xi$ has correlation
\begin{equation*}%\label{corr-noise}
  \langle\xi^\alpha(t) \xi^\beta(s)\rangle = 2k_B L^{\alpha\beta} \delta(t-s).
\end{equation*}
%\begin{equation}\label{exmp-projection}\left\{
%    \eqalign{
%      D^\alpha x &= k_B L^{\alpha\beta} \frac{\pt \log w}{\pt x^\beta} - k_B L^{\beta\gamma} \Gamma_{\beta\gamma}^\alpha, \\
%      Q^{\alpha\beta} x &= 2k_B L^{\alpha\beta},
%    }\right.
%\end{equation}

Comparing \eref{SDE} with \eref{generalized-Langevin}, we see that the Riemannian metric $(L^{\alpha\beta})$ is a generalization of the Onsager matrix to inhomogeneous transport processes (as $(L^{\alpha\beta})$ can vary for different thermodynamic variables), and the function $S$ in \eref{generating-func} provides a definition of entropy for nonequilibrium systems. The relation \eref{action-1} between action functional $\mathcal S$ and entropy $S$ suggests that the quantity $S(t,x(t))$ can be regarded as a trajectory-dependent entropy \cite{Sei05} and $\mathcal S$ is the totality of change of entropy.
The relations \eref{momenta} and \eref{force} indicate that the generalized force $(F_\alpha)$ is just the conjugate momentum $(p_\alpha)$.

It follows from the Legendre transform \eref{Legendre} that $D_\nabla^\alpha x = L_{\alpha\beta} p_\alpha$ and the Lagrangian associated to $H$ of \eref{exmp-Hamiltonian} is given by
\begin{equation*}
  L_0(x, D^{}_\nabla x) = \frac{1}{2} L_{\alpha\beta}(x) D^\alpha_\nabla x D^\beta_\nabla x = \frac{1}{2} L^{\alpha\beta}(x) p_\alpha p_\beta.
\end{equation*}
This lead to the following entropy production formula, by \eref{DS}:
\begin{equation*}
  \frac{d\langle S\rangle}{dt} = \langle L_0\rangle = \frac{1}{2} \langle L^{\alpha\beta} p_\alpha p_\beta \rangle,
\end{equation*}
which recovers Onsager's formula \eref{ent-prod-1}.

\section{Time reversibility}\label{sec-5}

We are going to describe the relations between the traditional (equilibrium) version of reversibility, due to Kolmogorov, and the version associated with Schr\"odinger's one, summarized briefly in Section \ref{sec-Schrodinger}. Their condition of validity and some consequences will be mentioned as well.

\subsection{Kolmogorov's reversibility vs. Schr\"odinger's reversibility}

Let us consider a Langevin equation of a form more general than \eref{generalized-Langevin}, but now interpreted in a phase space,
\begin{equation}\label{Langevin}
\left\{\eqalign{
  dx = v dt, \\
  dv = -\nabla U(x)dt - \gamma v dt + \sqrt{2\gamma k_B T} dW,
}\right.
\end{equation}
where $U:\R^n\to \R$ is a smooth potential such that $e^{-U(x)} \in L^1(\R^n)$, $\gamma$ a positive constant describing the damping strength, $W$ is a standard $n$-dimensional Brownian motion. Written formally as an ordinary differential equation, it is generally regarded as a Newton equation with a dissipative term proportional to the velocity, and a source of noise
\begin{equation*}
  \ddot x = -\nabla U(x) - \gamma \dot x + \sqrt{2\gamma k_B T} \dot W,
\end{equation*}
whose ``acceleration'', given definition \eref{Langevin}, is particularly problematic (the last term of r.h.s. being the white noise). The process $(x,v)$ is Markovian and hypoelliptic (there is no noise in the first equation of \eref{Langevin}). Moreover, this process is ergodic, with a (Gibbs) product invariant measure
\begin{equation*}
  \mu(dx,dv) = Z^{-1} \exp\left\{-\frac{H_0(x,v)}{k_B T}\right\} dxdv,
\end{equation*}
where $H_0$ is the Hamiltonian $\frac{1}{2}|v|^2 + U(x)$, and whose normalization constant $Z$ is the partition function on $\R^{2n}$. The process $(x,v)$ is non-reversible in Kolmogorov equilibrium sense \cite{Kol92} (which we shall recall below), i.e., its generator $\L$ is not self-adjoint in $L^2(M)$. The FP equation of the Langevin equation \eref{Langevin} takes the form $\frac{\pt p}{\pt t} = \L^* p$, for $\rho = \rho_t(x,v)$, with
\begin{equation*}
  \L^* \rho = -v \cdot \nabla_x \rho + \nabla_x U \cdot \nabla_v \rho + \gamma \nabla_v (v \rho) + \gamma k_B T \Delta_v \rho,
\end{equation*}
the existence and uniqueness of solution of this equation are known, as well as its rate of convergence to equilibrium, but the proofs are difficult \cite{Vil09}. Moreover, as can be expected, those equations cannot be solved explicitly, except for quadratic potentials $U$.

It is relevant here, to recall that this analytical notion of reversibility \cite{Kol92} was expressed originally for a single stationary diffusion process with invariant measure $\mu(dx) = \rho(x)dx$. The forward and backward transition probability densities, respectively $p(x,t,dy)$ and $p_*(dx,t,y)$, satisfy the following ``detailed balance condition'':
\begin{equation}\label{detailed-balance}
  \rho(x) p(x,t,dy) dx = p_*(dx,t,y) \rho(y) dy,
\end{equation}
where $p$ solves the usual Fokker-Planck (FP) equation
\begin{equation}\label{FP}
  \frac{\pt p}{\pt t} = -\vf{y^i} (b^i(y) p) + \frac{1}{2} \frac{\pt^2}{\pt y^i \pt y^j} (a^{ij}(y) p),
\end{equation}
with drift vector field $(b^i)$ and positive definite diffusion tensor $(a^{ij})$, and $p_*$ denotes the probability transition to come back around $x$ from a future location $y$, during the time interval $[0,t]$. Kolmogorov formulated the reversibility as the following equation
\begin{equation*}
  p_*(dx,t,y) = p(y,t,dx),
\end{equation*}
and proved that this equation holds if and only if the drift is a gradient:
\begin{equation}\label{Kolm}
  b^i = \frac{a^{ij}}{2} \frac{\pt (\log\rho)}{\pt x^j}. %= \frac{1}{2} \nabla\log\rho,
\end{equation}
%where the gradient $\nabla$ is with respect to the diffusion tensor $a$ as a Riemannian metric.
This stationary gradient form of the drift \eref{Kolm}, involved in his equilibrium reversibility condition \cite{Kol92} was therefore generalized via forward and backward transition probabilities \eref{forward-ts}, \eref{backward-ts} to associated drifts describing Schr\"odinger's time-inhomogeneous processes. Given an initial probability density $\rho$, the future one follows by integration w.r.t. the elementary solution of \Eref{FP}. This means that it is sufficient to solve an initial value FP problem for $\rho$. The detailed balance condition \eref{detailed-balance} expresses two equivalent ways to define the two-times joint probability of a reversible homogeneous Markov process.
%Integrating each side of \eref{detailed-balance}, we can also get two equivalent expressions for its equilibrium marginal distribution:
%\begin{equation*}
%  \mu(dy) = \rho(x) p(x,t,dy) dx = p_*(dy,t,x) \rho(x) dx.
%\end{equation*}

Comparing relation \eref{detailed-balance} with \eref{noneq-balance}, it is clear that Schr\"odinger's nonequilibrium notion of reversibility, needed for his variational problem, is the general form (but local in time) of Kolmogorov's one, formulated five years earlier. As said in the introduction, Schr\"odinger's context was, apparently, the one of classical statistical physics but his problem was essentially inspired by his doubts about the foundations of quantum mechanics. It is only when the unexpected relation between Nelson's stochastic mechanics \cite{Nel01} (a radical attempt to interpret the quantum wave equation itself as a local in space, regularized, version of Newtonian dynamics) and Schr\"odinger's 1931 idea was discovered \cite{Zam86} that Kolmogorov's conditions \eref{detailed-balance} and \eref{Kolm} were suitably generalized (cf. \cite{Zam86,Zam15}).
%Also notice that the description of a Markov process in term of a transition probability, like \Eref{forward-ts}, introduces necessarily an arrow of time (i.e., is ``irreversible'').
The special feature of Schr\"odinger(-Bernstein) diffusions is that they admit simultaneously two transition probabilities, allowing to preserve his nonequilibrium notions of detailed balance and reversibility.

We shall give a theorem summarizing the key conditions for the construction of the diffusions in $\R^n$ solving Schr\"odinger's variational problem for the special class of Hamiltonian $\mathcal H$, as in \eref{backward-heat}--\eref{forward-heat}, of the form
\begin{equation*}
  \mathcal H= \mathcal H_0 + U = -\frac{\hbar^2}{2} \Delta + U(x),
\end{equation*}
with $U$ an additional scalar potential. The positivity of the above free integral kernel $h_0$ is obvious so the one of $h(x, t-s, y) = (e^{-(t-s) \mathcal H} \delta_x)(y)$ is a condition on the potential $U$. The theorem involves results of a number of authors, along the years, following Schr\"odinger's seminal idea (cf. \cite{AYZ89,Beu60,CZ91} for more).

\begin{theorem}\label{Schrodinger}
  Let $\mathcal H= \mathcal H_0 +U$ densely defined in $L^2(\R^n)$, lower bounded self-adjoint Hamiltonian, with $U:\R^n\to\R$ a real measurable function in the Kato class $\mathcal K^n$ \cite{AYZ89}. For instance, when $n=3$, the harmonic oscillator and Coulomb potential are in $\mathcal K^3$. If $\max(-U,0)\in \mathcal K^n$ and $U \ind_{B(0,L)} \in \mathcal K^n$, the kernel $h(x,t-s,y)$, $t\ge s$ is jointly continuous in the $3$ variables and nonnegative. Let $\rho_{t_0}(x)$ and $\rho_{t_1}(y)$ be two strictly positive probability densities, for $\mathcal H$ as before. Then positive, not necessarily integrable, solution $(\psi,\phi)$ of \Eref{Sch-system} for $h$ as before, exists and is unique. The optimal diffusion $X(t)$ is, therefore, completely determined.
\end{theorem}

Schr\"odinger himself checked, in \cite{Sch32}, that the simplest diffusion on $\R$ associated with $\mathcal H_0$ (with zero potential $U\equiv 0$), the one-dimensional Brownian motion $X(t) = x+ W(t-s)$, starting from $x$ at time $s$ was compatible with his approach, though as a degenerate, manifestly nonequilibrium example.

In this case we are given that $\rho_{t_0}(x) = \delta_x$ and $\rho_{t_1}(y) = [2\pi (t_1-t_0)]^{-\frac{1}{2}} e^{-\frac{|y-x|^2}{2(t_1-t_0)}}$ and the solution of \Eref{Sch-system} is the trivial one $\phi(y) = 1$, $\psi(x) = \delta_x$. The corresponding solutions of \Eref{backward-heat} and \Eref{forward-heat}, $t\in[t_0,t_1]$ are $w(x,t) \equiv 1$ and $w^*(y,t) = [2\pi (t-t_0)]^{-\frac{1}{2}} e^{-\frac{|y-x|^2}{2(t-t_0)}}$. Using \eref{forward-ts} and \eref{backward-ts} it is easy to verify that the usual (forward) drift $B$ is indeed the conditional expectation
\begin{equation*}
  B(x,t) = \lim_{\Delta t\to0^+} \E\left[ \frac{X(t+\Delta t)-X(t)}{\Delta t} \bigg| X(t) = x \right] = \frac{\nabla w}{w} (x,t) = 0,
\end{equation*}
and the backward one
\begin{equation*}
  B_*(y,t) = \lim_{\Delta t\to0^+} \E\left[ \frac{X(t)-X(t-\Delta t)}{\Delta t} \bigg| X(t) = y \right] = -\frac{\nabla w^*}{w^*} (y,t) = \frac{y-x}{t-t_0},
\end{equation*}
whose singularity in $t=t_0$ expresses our singular initial condition $X(t_0) = x$. Also notice that $\phi$, here, is not integrable. Of course, given $\mathcal H=\mathcal H_0 + U$, each new data of $(\rho_{t_0}, \rho_{t_1})$ as in Theorem \ref{Schrodinger}, will be associated to a new diffusion. This is the origin of the rich dynamical content of the stochastic theory described here.

In this elementary example, the Hamiltonian function of system \eref{canonical-stoch-Hamilton-eqns} reduces, by \eref{canonical-H} in this flat one-dimensional case, to
\begin{equation*}
  H(x,p,o) = H_0(x,p) + \frac{1}{2} o,
\end{equation*}
where $o = \frac{\pt p}{\pt x}$ and $H_0 = \frac{1}{2} |p|^2$ is the free classical (but Euclidean) Hamiltonian.
%$H$ is proportional to
%\begin{equation*}
%  H(x,p,o) = \frac{1}{2} |p|^2 + \frac{1}{2} \frac{\pt p}{\pt x}.
%\end{equation*}
In other words, the logarithm substitution $S=\log w$, for $w$ solving \Eref{forward-heat}, provides HJB with final boundary condition (since $p=\nabla S$),
\begin{equation*}
  \frac{\pt S}{\pt t} + H_0(x,\nabla S) + \frac{1}{2} \Delta S = 0.
\end{equation*}
However, this solution is trivial because of our choice $\phi = w=1$ for \Eref{forward-heat}. The backward HJB counterpart of \eref{HJB-Riem}, namely,
\begin{equation*}
  \frac{\pt S^*}{\pt t} + H_0(x,-\nabla S^*) - \frac{1}{2} \Delta S^* = 0
\end{equation*}
related with \Eref{backward-heat} by the transformation $S^* = -\log w^*$ and $p=\nabla S^*$, is more interesting since $w^*$ is the elementary solution of the Cauchy problem \Eref{backward-heat}. The resulting drift is $B_*$ founded before.

In its traditional (forward) description, with zero drift, the Brownian motion looks hardly ``reversible''. However, as a reciprocal process, its backward drift displays its singular initial boundary condition, an observation known for a long time (cf. for instance, \cite[Example 8.3]{Wen81}), but rarely used in applied sciences.

The Lagrangian $L_0$ associated with $H_0$ is the free one $L_0(X(t), DX(t)) = \frac{1}{2}|DX(t)|^2$ which, for the Brownian motion, corresponds to a trivial solution of the stochastic Euler-Lagrange equation \eref{EL}, namely, $DDX(t)=0$.
This example provides us the motivation to elaborate on the classical notion of time reversal and Schr\"odinger's one.

Traditionally, for any classical observable in phase space $O(x(t),p(t))$, a time reversal operator $T_c$ is introduced so that $\mathcal T_c(x(t),p(t))= (x(t),-p(t))$. Generally, the classical Hamiltonian observable $H_0(x,p)$ of the system is invariant under $\mathcal T_c$. Then the microscopic reversibility of the system means that if $(x(t),p(t))$ are solutions of Hamilton's equations with initial condition $(x(0),p(0))$, so are $\mathcal T_c(x(t),p(t))$, with time reversed boundary condition $\mathcal T_c(x(0),p(0))$.

For Bernstein's reciprocal process $X(t)$ solving Schr\"odinger's problem, the notion is a bit more sophisticated, because of the non-differentiability of their paths. Let us summarize it for the special class of Hamiltonian operators $\mathcal H$ of Theorem \ref{Schrodinger}. %quantizations of classical ones $h(x(t),p(t))$, whose momenta reduce to time derivatives.

As already said, it follows from the special form \eref{Bonn-1} of the probability density looked for by Schr\"odinger that those process $X(t)$ are themselves invariant under time reversal. Regarding the drifts in the simplest example of Brownian motion following Theorem \ref{Schrodinger}, we introduce the regularized forward time derivative as limit of conditional expectation of difference, as in \eref{mean-d-x},
\begin{equation*}
  DX(t) = \lim_{\Delta t\to0^+} \E\left[ \frac{X(t+\Delta t)-X(t)}{\Delta t} \bigg| X(t) \right].
\end{equation*}
Let us compute this drift for the time reversal $\overleftarrow X(t) = X(-t)$, we find $D\overleftarrow X(t) = -D_*X(-t)$, where $D_*$, the backward derivatve is now defined by,
\begin{equation*}
  D_*X(t) = \lim_{\Delta t\to0^+} \E\left[ \frac{X(t) - X(t+\Delta t)}{\Delta t} \bigg| X(t) \right],
\end{equation*}
generally distinct from $DX(t)$, as illustrated by the above example. This means that our counterpart of the classical time reversal operator $\mathcal T_c$ becomes
\begin{equation*}
  \mathcal T(X(t),DX(t)) = (X(t), - D_*X(t)).
\end{equation*}

So we can indeed make all computations using traditional tools of (forward) It\^o's calculus, but be aware that their results have a backward version and that fully dynamical laws have to be invariant under $\mathcal T$. For instance, our stochastic Hamiltonian equations \eref{stoch-Hamilton-eqns} have a backward counterpart, needed to re-establish the reversibility of our stochastic dynamical system. So, Schr\"odinger's reversibility, encapsulated in his nonequilibrium detailed balance condition \eref{noneq-balance}, is a consequence of the stochastic deformation of classical microscopic reversibility, founded on the $\mathcal T_c$ operator.

For instance, coming back to our elementary Brownian case and using the stochastic time reversal $\mathcal T$, the backward version of the free Lagrangian $L_0$ is $L_0(t, X(t), D_*X(t)) = \frac{1}{2} |D_*X(t)|^2$. The backward version of the free Euler-Lagrange equation becomes
\begin{equation*}
  \eqalign{
    D_* D_* X(t) &= D_* B_* = D_* \left( \frac{X(t) - x}{t-s} \right) \\
    &= \left( \frac{\pt}{\pt t} + \frac{X(t) - x}{t-s} \nabla_y - \frac{1}{2}\Delta_y \right) \left( \frac{y - x}{t-s} \right)\bigg|_{y=X(t)} = 0.
  }
\end{equation*}
Notice that the minus sign in front of the Laplacian term comes from backward It\^o's calculus (cf. \cite{Zam15} and reference therein).

To conclude this part, we observe that, motivated by the product form of the density in \Eref{Bonn}, Schr\"odinger was interested in Markovian processes. A more general construction, due to B. Jamison, started from any joint probability $\mu_{t_0,t_1}(dx,dy)$, not necessarily of the form \eref{Markov-marginal} (\Eref{Markov-marginal} is the only form producing Markovian processes, cf. Section \ref{sec-2-2}). In general, we call Bernstein's reciprocal processes the processes following from an arbitrary choice of joint probability. Those processes have been proved to be relevant as counterparts of quantum statistical systems \cite{VZ16}.

More information and examples can be found in \cite{CZ03,Zam15}.

\subsection{Schr\"odinger's approach as an Euclidean quantum mechanics}

As another illustration of the key relevance of this aspect in the stochastic dynamical theory inspired by Schr\"odinger, let us mention that the rigorous (Euclidean) versions of Feynman's path integral interpretation of Heisenberg commutation relation requires this time reversal of the drift (cf. \cite{Zam15}).

Schr\"odinger's reinterpretation (by anticipation!) of Feynman's path integral commutation relations highlights the deep analogies between his stochastic dynamical approach and quantum mechanics of pure states. They are rooted in the fundamental role of the two adjoint heat equations \eref{backward-heat} and \eref{forward-heat}, and the Euclidean Born's interpretation of \eref{Bonn}. Let us summarize some key relations of the resulting Schr\"odinger's Euclidean Quantum Mechanics, namely between the Markovian processes solving his stochastic dynamics and operators calculus founded on the class of Hamiltonian operators of Theorem \ref{Schrodinger}.

Let $f(X(t),t)$ be a smooth function of the Markovian process of Theorem \ref{Schrodinger}. Then its regularized forward derivative (or infinitesimal generator), as in \eref{mean-d}, can be expressed, as a differential operator, by
\begin{equation*}
  Df(X(t),t) = \frac{1}{w(X(t),t)} \left( \frac{\pt}{\pt t} - \mathcal H \right) (fw)(X(t),t).
\end{equation*}
The proof is an exercise of It\^o's calculus using the form of $\mathcal H$ in Theorem \ref{Schrodinger} and the fact that $w$ solves \Eref{backward-heat}. Explicitly, the l.h.s. describes the dynamics of $X(t)$ and the r.h.s. has close relation with the Hamiltonian operator $\mathcal H$, generator of Euclidean dynamics, for a given Euclidean ``state'' $w$.

The backward version of $Df(X(t),t)$ results from the time reversed expression, for $w^*$ solving \Eref{backward-heat},
\begin{equation*}
  D_*f(X(t),t) = \frac{1}{w^*(X(t),t)} \left( \frac{\pt}{\pt t} + \mathcal H^* \right) (fw^*)(X(t),t),
\end{equation*}
where $\mathcal H^* = \mathcal H$ as $\mathcal H$ is self-adjoint.

It is possible to construct a Hilbert space approach of Schr\"odinger's Euclidean dynamics, describing how to translate properties of the process $X(t)$ in terms of an operators calculus, in analogy with the quantum model \cite{AYZ89}. In it, a number of purely informal results of Feynman's path integral method make perfect mathematical sense (Heisenberg's commutation relations being only one of them).

The expectation of (Euclidean) observables in Hilbert space take a familiar form. For instance, the one of the momentum is
\begin{equation*}
  \langle P \rangle_w = \int w^* P^+ w dx = \int ww^* \left( \frac{P^+ w}{w} \right) dx,
\end{equation*}
for $P^+ = \nabla$ so that the forward drift is, indeed $B=\frac{\nabla w}{w}$. Since $P^+$ is not symmetric, the backward one is $\int ww^* \left( \frac{P w^*}{w^*} \right) dx$ namely $B^* = -\frac{\nabla w^*}{w^*}$.

Regarding the energy observable, the Hamiltonian $\mathcal H$ of Theorem \ref{Schrodinger} is symmetric and
\begin{equation*}
  \langle \mathcal H \rangle_w = \int w^* \mathcal H w dx = \int ww^* \left( \frac{\mathcal H w}{w} \right) dx,
\end{equation*}
namely, after using $\frac{\Delta w}{w} = |\frac{\nabla w}{w}|^2 + \nabla \cdot \left( \frac{\nabla w}{w}\right)$, we find the random variable $\frac{1}{2}|B|^2 + \frac{1}{2}\nabla\cdot B - U$, where the sign change of the potential is usual in the Euclidean world and the extra divergence term comes from It\^o's deformation or, equivalently, the second-order term of HJB equation \eref{HJB-Riem}. Indeed, in this Euclidean case, $B=\nabla S$, $H_0 = \frac{1}{2}|p|^2 - U(x)$ and the second-order correction $\frac{1}{2} g^{jk} (o_{jk} - p_i \Gamma_{jk}^i)$ reduces to $\tr o$.

\section{Conclusion and prospects}

Considering the evolutions of idea between statistical and quantum physics since the old formulation by Schr\"odinger of his variational problem, the status of the stochastic dynamical theory founded on its solution, whose Hamiltonian side is summarized here, may appear paradoxical.

On one side, it is a classical analogy with quantum mechanics, and was designed as such by Schr\"odinger. He was not only anticipating the ``Euclidean revolution'' in QFT (Schwinger, Symanzik, Nelson, etc), where time becomes purely imaginary and Feynman's path integrals become rigorous, but as well the multiple attempts to make sense of the (still missing) probabilistic content of elementary quantum theory for pure states.

A curious consequence of the resulting stochastic dynamical theory, although perfectly classical in appearance (except for boundary conditions) is that its associated processes manifest indeed striking similarities with the mathematically nonexistent ones introduced by Feynman, including properties generally regarded as restricted to the quantum domain. This was certainly one of the initial motivations of Schr\"odinger, often sarcastic about foundational issues of this theory.

The role of time reversal lies at the heart of Schr\"odinger's analogy, which may seem strange in the context of nonequilibrium statistical mechanics. But consider action functionals of a system associated with an Hamiltonian operator on $\R^n$ in the class of Theorem \ref{Schrodinger}. It is known \cite{Zam86,CZ91} that there are two such conditional expectations at a given time $t$ inside $[t_0,t_1]$, one defined in the interval $[t_0,t]$ the other in $[t,t_1]$. The first one involves a regularized (Euclidean) Lagrangian $\frac{1}{2}|DX|^2 -U(X)$ and the other $\frac{1}{2}|D_*X|^2 -U(X)$ (they are time reversed of each other). Now compute the absolute expectation, with density \eref{Bonn-1}, of those functionals on $[t_0,t_1]$. Using \eref{action-1} and its backward version, one finds that
\begin{equation*}
  \eqalign{
    &\ \E[\log \rho(X(t_1),t_1) - \log \rho(X(t_0),t_0)] \\
    =&\ \frac{1}{2} \left[ \E\int_{t_0}^{t_1} \left( \frac{1}{2}|DX(t)|^2 +U(X(t)) \right) dt - \E\int_{t_0}^{t_1} \left( \frac{1}{2}|D_*X(t)|^2 +U(X(t)) \right) dt \right].
  }
\end{equation*}
So, the difference of expectations of our forward and backward actions coincides with a difference of Boltzmann entropies of diffusion $X(t)$ (\cite[p 59]{CWZ00}), an intriguing interplay between the statistical physics content of Schr\"odinger inspired theory and its (classical or quantum) mechanical one. The important role of time reversal in nonequilibrium systems has already been recognized a long time ago \cite{Gas13}. A book published quite recently by the same author certainly elaborates this aspect \cite{Gas22}.

There are a lot of new such relations waiting to be discovered in Schr\"odinger's Euclidean approach. To start with, there are a lot of notions of entropies going around in this framework, which should be systematically organized. Some of its above mentioned paradoxical aspects have to do with the local (in time) character of the detailed balance condition \eref{noneq-balance}. It is due to the fact that the starting time interval $I=[t_0,t_1]$ of construction of the processes is arbitrary but, in general, finite. How can this be compatible with the time asymptotic study of stationary, equilibrium macrostates of the system? The difficulty seems insuperable when computations involve retrograde heat equation like \eref{forward-heat}. But, as for Schr\"odinger's problem itself, the answer could be in the way to ask the question. For instance, a theorem of Robbins and Sigmund (1973), cited in \cite{KS91}, shows that a class of positive solutions $w$ of \Eref{forward-heat} has indeed, an infinite lifetime. They take the form of an Euclidean version of plane wave representation of solution of free Schr\"odinger's equation. The associated Berstein's Markovian diffusions should be natural candidates for time asymptotic study of such stochastic dynamical systems.

As observed before, Schr\"odinger's main interest was in Markovian processes. But Jamison's construction allows to start from any joint probability $\mu_{t_0,t_1}(dx,dy)$ of the boundary marginals. Examples of such Bernstein's reciprocal but non-Markovian processes have been considered \cite{VZ16} and should be of interest in physics. %They break, of course, any notion of reversibility.

In conclusion, we share Schr\"odinger's view, recalled in our introduction, that the validity of the notion of nonequilibrium detailed balance underlying his idea is not accidental but we can add, almost 90 years after the publication of his variational problem, that this notion has the potential, in our opinion, to transform some discussions around the foundations of nonequilibrium statistical physics.

\vspace{3mm}
\paragraph{Acknowledgements.}
It is a pleasure to thank the organizer of ``Koopman Methods in Classical and Quantum-Classical Mechanics'' held in July 2021, namely, Profs. D.I. Bondar, I. Burghardt, F. Gay-Balmaz, I. Mezic, C. Tronci. It was, unfortunately, an online seminar so that we could not enjoy the hospitality of the Wilhelm and Else Heraeus Foundation. But we look forward to do it in the future, since the scientific community involved is quite interesting.
%We would like to thank Prof.~XXX for helpful suggestions.
%We would like to thank the reviewers for their thoughtful comments and efforts towards improving our manuscript.
This paper is supported by FCT, Portugal, project PTDC/MAT-STA/28812/2017, ``Schr\"odinger's problem and optimal transport: a multidisciplinary perspective (Schr\"oMoka)''.

\bibliographystyle{unsrt}
\bibliography{StochHamRev-ref}

\end{document}